\documentclass{article}
\usepackage{spconf,amsmath,graphicx}

\usepackage{graphicx,verbatim}
\newcommand{\keypoint}[1]{\noindent\textbf{#1}\quad}
\usepackage{colortbl}
\usepackage{algorithm}
\usepackage{algpseudocode}
\usepackage[breaklinks,colorlinks]{hyperref} 

\usepackage{amsmath}
\usepackage{amsfonts}
\usepackage{color}
\definecolor{LightCyan}{rgb}{0.95,0.95,0.95}
\definecolor{LightGrey}{rgb}{0.85,0.91,0.98}
\usepackage{multirow}
\usepackage{booktabs}
\usepackage{enumitem}
\setlist{nosep, leftmargin=14pt}

\usepackage[most]{tcolorbox}
\tcbset{
  highlightbox/.style={
    enhanced,
    colback=yellow!20,
    colframe=yellow!20,
    arc=6pt,
    boxrule=0pt,
    left=6pt, right=6pt, top=4pt, bottom=4pt,
    width=\linewidth,
    sharp corners=downhill,
  }
}

\usepackage{mwe} 

\title{On The Robustness of Foundational 3D Medical Image Segmentation Models Against Imprecise Visual Prompts}
\name{Soumitri Chattopadhyay, Ba\c{s}ar Demir, Marc Niethammer}
\address{University of California, San Diego}

\begin{document}
%
\maketitle
\begin{abstract}
While 3D foundational models have shown promise for promptable segmentation of medical volumes, their robustness to imprecise prompts remains under-explored. In this work, we aim to address this gap by systematically studying the effect of various controlled perturbations of dense visual prompts, that closely mimic real-world imprecision. By conducting experiments with two recent foundational models on a multi-organ abdominal segmentation task, we reveal several facets of promptable medical segmentation, especially pertaining to reliance on visual shape and spatial cues, and the extent of resilience of models towards certain perturbations. Codes are available at: \href{https://github.com/ucsdbiag/Prompt-Robustness-MedSegFMs}{https://github.com/ucsdbiag/Prompt-Robustness-MedSegFMs}



\end{abstract}
\section{Introduction}


Interactive, or \textit{promptable}, segmentation models~\cite{sam,sam_survey,sammed3d,segvol,nninteractive} segment user-specified regions of interest, enabling general foreground--background separation beyond class-specific training. These models rely on \textbf{\textit{visual prompts}} that are localized spatial cues that indicate the target structure. For natural images, SAM~\cite{sam} supports point clicks, bounding boxes, and masks. Medical imaging adaptations extend these ideas: MedSAM~\cite{medsam} and SAM-Med2D~\cite{sam_med2d} use bounding boxes in 2D, while SAM-Med3D~\cite{sammed3d}, SegVol~\cite{segvol}, and nnInteractive~\cite{nninteractive} employ clicks or combinations of multiple prompts (clicks, boxes, scribbles etc.) for 3D volumes. Many of these models refine predictions iteratively, using the previous segmentation as a \textbf{\textit{dense mask prompt}} to guide subsequent updates~\cite{sammed3d,nninteractive}.

Despite these advances in promptable 3D medical image segmentation, performance of these models is \textit{highly dependent on the \underline{quality of input prompts}}. In practice, obtaining pixel-perfect visual prompts is often unrealistic: due to time constraints of high-quality annotation, lack of expert annotators, or atlas-driven prompt generation from imperfect registration \cite{unigradicon, mmregseg}. Yet, robustness to imprecise prompts is largely unexplored, despite growing clinical adoption.


In this work, we conduct a systematic probing of prompt robustness for 3D medical segmentation foundation models. For this study, we focus on \textbf{\textit{dense visual prompts}} rather than manual sparse interactions (clicks, boxes, etc.), as they are \textit{more flexible}: they enable automated prompt generation from auxillary segmentation~\cite{nnunet}, registration~\cite{unigradicon,mmregseg,mmgradicon} or supervoxels~\cite{vista3d,felzenszwalb2004efficient}, represent the iterative refinement paradigm where \textit{predictions become prompts}~\cite{sammed3d,nninteractive}, as well as can be used to simulate sparse interactions. From gold standard masks, we simulate realistic imprecise dense visual prompts through \textbf{\textit{morphological perturbations}} and \textbf{\textit{spatial translations}}. Leveraging perturbed prompts, we experiment with two state-of-the-art foundational models, namely nnInteractive~\cite{nninteractive} and SAM-Med3D~\cite{sammed3d}, on a multi-organ abdominal segmentation task~\cite{btcv}. Our controlled oracle experiments reveal \textbf{\textit{(i)}} how models respond to varying prompt perturbations; \textbf{\textit{(ii)}} how organ geometries relate to imprecise visual prompting; \textbf{\textit{(iii)}} traits about their reliance on shape or boundary-aware priors compared to dense voxel-level precision.


\vspace{0.1cm}
\noindent To our knowledge, \textit{this is the first robustness analysis of 3D medical segmentation foundation models against imprecise visual prompts.} While our experiments are synthetic, they mimic real-world scenarios where pixel-perfect prompts are unrealistic, making robustness evaluation timely and relevant.



\section{Materials and Methods}\label{sec:materials}

\keypoint{Models:}We choose two recently proposed state-of-the-art foundational models for 3D medical segmentation -- \textbf{nnInteractive}~\cite{nninteractive} and \textbf{SAM-Med3D}~\cite{sammed3d}. While the latter expands the SAM~\cite{sam} model architecture to 3D volumes, nnInteractive uses a CNN backbone with prompts encoded in different channels of the input (along with the image). 

\vspace{0.2cm}

\keypoint{Designing imprecise visual prompts:}First, we perform morphological corruptions: \textbf{\textit{(i)}} \textbf{dilation} and \textbf{\textit{(ii)}} \textbf{erosion}, on the gold standard dense masks to yield perturbed coarse masks to be used as dense prompts. We vary the \underline{radius} of dilation and erosion as integers in range $[1, 8]$ to obtain varying degrees of prompt coarseness. Dilated and eroded regions of interest typically mimic practical scenarios where voxel-level precision is often compromised for faster processing, lack of domain expertise, or automatically generated prompts e.g. by registration~\cite{unigradicon} from an existing annotated image. Formally,

\vspace{-0.5cm}
\begin{equation}
\begin{aligned}
    \mathcal{M}_{\texttt{dilate}} &= \mathcal{M}_{\texttt{orig}} \oplus \mathcal{B}_r, \quad
    \mathcal{M}_{\texttt{erode}} = \mathcal{M}_{\texttt{orig}} \ominus \mathcal{B}_r 
\end{aligned}
\end{equation}

\noindent where $\oplus$ and $\ominus$ denote morphological dilation and erosion operations; $\mathcal{B}_r$ expands/contracts by $r$ voxels in all directions, $r \in \{1, \ldots, 8\}$.

Furthermore, to disentangle the roles of boundary localization and dense-level context, we also include \textbf{\textit{(iii)}} \textbf{boundary-preserving erosion}, a synthetic corruption that maintains the boundary outline of the mask while forming a cavity within its interior. This contrasts with standard erosion described above (which loses boundary information), but in turn lets us analyze the effect of shape-aware visual cues for promptable models. Formally, for boundary thickness $d \in \{1, \ldots, 8\}$, we have

\vspace{-0.3cm}
\begin{equation}
\begin{aligned}
    \mathcal{M}_{\texttt{cavity}} &= \mathcal{M}_{\texttt{orig}} \setminus (\mathcal{M}_{\texttt{orig}} \ominus \mathcal{B}_d)\,.
\end{aligned}
\end{equation}

Additionally, we also probe with \textbf{\textit{(iv)}} \textbf{laterally translated masks} i.e. dense visual prompts which are spatially shifted from the true target region, again mimicking practical clinical usage when prompts are automated from unaligned images~\cite{unigradicon}. We apply small magnitudes of translation on all of the axes (without any morphological changes), since there is a large variation in sizes of organs and uncontrolled shifts may completely lose the region of interest in context. 

\vspace{-0.4cm}
\begin{equation}
\begin{aligned}
    \mathcal{M}_{\texttt{shift}}[x,y,z] = \mathcal{M}_{orig}[x+\delta_x, y+\delta_y, z+\delta_z]\\
    \text{where } \delta_{\{x,y,z\}} \sim \mathcal{U}\{-3, \ldots, 3\}\,.
\end{aligned}
\end{equation}

 Having obtained the perturbed dense visual prompts, we follow the respective model workflows to use them as inputs and obtain segmentation predictions using the frozen models.

\section{Experimental Setup}\label{expts}

\keypoint{Implementation.}All source codes were implemented using PyTorch~\cite{pytorch}. We used the publicly available code repositories as well as pre-trained released checkpoints of the respective foundational models to develop our dense-promptable inferencing setup. For nnInteractive\footnote{\href{https://github.com/MIC-DKFZ/nnInteractive}{https://github.com/MIC-DKFZ/nnInteractive}}~\cite{nninteractive}, all pre- and post-processing is handled within the model itself, hence we simply feed in the volumes with the augmented masks as dense prompts, all in their original resolutions. For SAM-Med3D\footnote{\href{https://github.com/uni-medical/SAM-Med3D}{https://github.com/uni-medical/SAM-Med3D}}~\cite{sammed3d}, the input volumes were first resampled to $1.5mm\times1.5mm\times1.5mm$ spacing, ROI cropped at $128\times128\times128$ around the organs, and Z-score normalized, following their exact preprocessing steps~\cite{sammed3d}. We conducted our experiments on a 48GB Nvidia RTX A6000 GPU.

\vspace{0.2cm}
\keypoint{Testing Dataset.} We perform our evaluations on the Beyond The Cranial Vault \textbf{(BTCV)} Challenge~\cite{btcv}, a multi-organ segmentation dataset comprising $30$ abdominal CT volumes. The organs encompassed by the dataset include: \textit{liver, kidneys, spleen, gallbladder, esophagus, adrenal glands, stomach, aorta, pancreas, inferior vena cava}, and the \textit{portal and splenic vein} (i.e. $13$ structures). The rationale for choosing this dataset is two-fold: \textbf{\textit{(i)}} the \textit{wide range of sizes of organs} covered in this dataset allows for reliable and insightful probing with imprecise visual prompts; \textbf{\textit{(ii)}} this dataset was \textit{not part of the training} of either of these foundational model checkpoints\footnote{ nnInteractive ckpt. trained on: \href{https://www.codabench.org/competitions/5263/}{https://www.codabench.org/competitions/5263/}}.

\vspace{0.2cm}
\keypoint{Metric.}We report \textbf{Dice} scores, a de facto standard metric for evaluating segmentation models~\cite{dice, chattopadhyay2025zeroshot}.

\section{Findings and Analysis}

\vspace{-0.2cm}
\noindent We highlight the key findings in terms of trends, and follow them up with discussions in the subsequent sections.

\begin{tcolorbox}[highlightbox]
\textbf{Models show greater robustness if the visual prompts contain boundary information.}
\end{tcolorbox}


\begin{figure}
    \centering
    \includegraphics[width=1\linewidth]{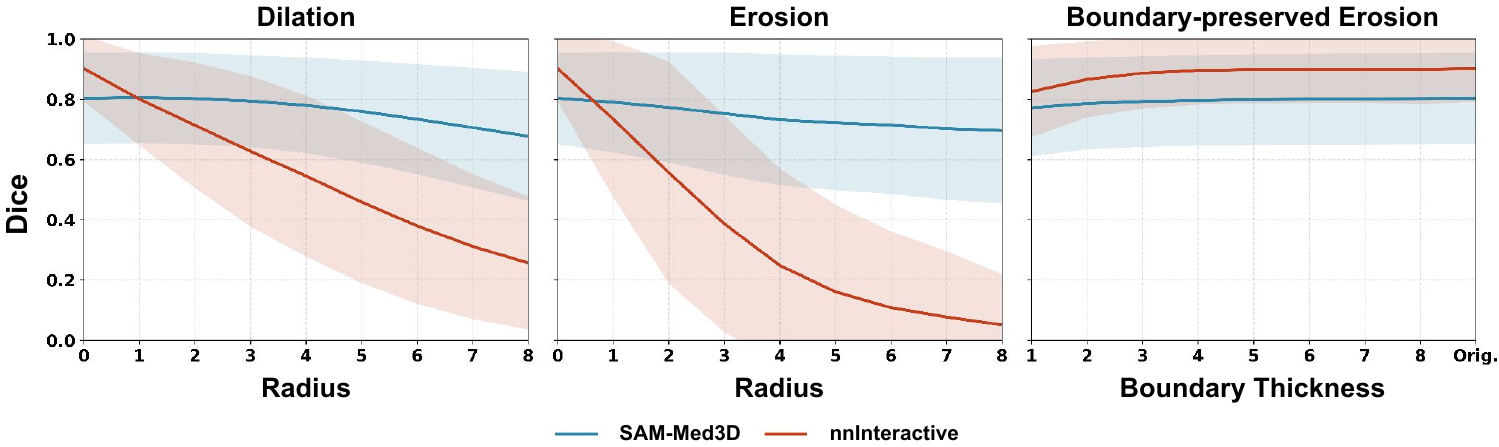}
    \vspace{-0.8cm}
    \caption{Mean trends aggregated over all organs for different prompt perturbations. The respective color bands denote standard deviations across all samples. Models show greater resilience to dilated prompts as compared to eroded ones, as well as to prompts containing bounding shape information.}
    \vspace{-0.25cm}
    \label{fig:global_trend}
\end{figure}

\begin{figure}
    \centering
    \includegraphics[width=1\linewidth]{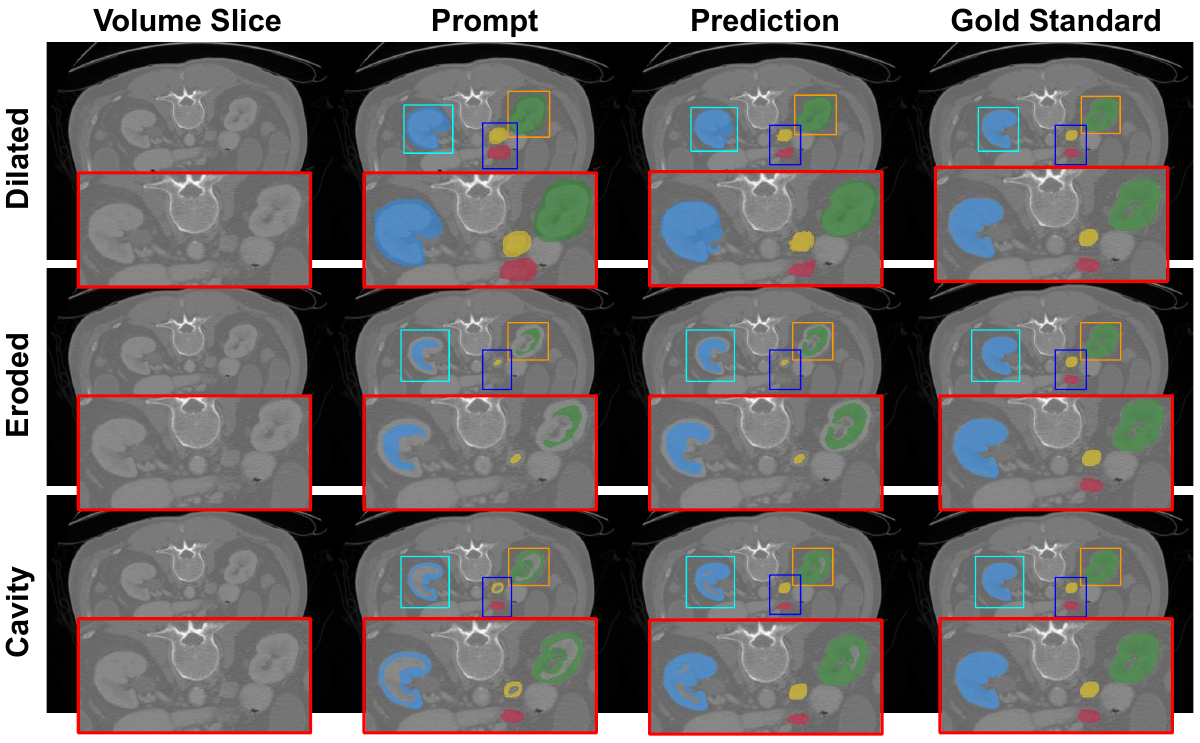}
    \vspace{-0.7cm}
    \caption{Qualitative results of nnInteractive for different prompt perturbations. Inset shows zoomed-in regions containing the organs, their perturbed prompts, predictions and the gold standard. As can be seen, the model over- and under-segments for dilated and eroded prompts, while it is more reliable when the prompt has bounding voxels (``Cavity'').}
    \vspace{-0.4cm}
    \label{fig:nnint_augs}
\end{figure}

In \autoref{fig:global_trend}, we show the aggregated global trends over all organs of BTCV, across varying strengths of perturbations -- dilation, erosion and boundary-preserved erosion. We observe that while segmentation performance deteriorates with increasing coarseness across all of the perturbations, models show greater resilience when the structural boundary is preserved in the visual prompt compared to the other cases. Both dilated and eroded masks coarsely indicate the target region but fail to provide explicit spatial constraints on segmentation extent, leading to suboptimal segmentation. Conversely, boundary-preserving erosion maintains shape topology, effectively ``telling the model'' the bounds within which to segment -- thereby injecting a stronger geometric prior and enabling greater robustness. We further provide qualitative examples to support this empirical observation in \autoref{fig:nnint_augs}, where clearly nnInteractive follows the dilated or eroded mask prompts and ends up segmenting beyond the true boundaries of structures, while it ``fills up'' the cavity of a boundary-informed visual prompt. Critically, boundary-preserving and fully eroded masks can have identical corruption levels (percentage of pixels removed), yet models show greater robustness to the former, isolating the effect of preserved spatial geometry over dense voxel-level precision.

It should be noted that boundary-preserving erosion is a less-realistic simulation of imprecise prompting compared to other corruptions in clinical settings. However, we justify its inclusion given that we focus on synthetic simulations, and the empirical findings reveal a salient trait of promptable segmentation models  -- \textbf{\textit{they benefit from shape descriptive visual prompts}} (since a boundary-preserving dense prompt provides a stronger shape prior compared to dilated/eroded ones).

\begin{tcolorbox}[highlightbox]
    \textbf{Segmentation robustness to imprecise prompts varies greatly with geometry of the target structures.}
\end{tcolorbox}

\begin{figure}
    \centering
    \includegraphics[width=1\linewidth]{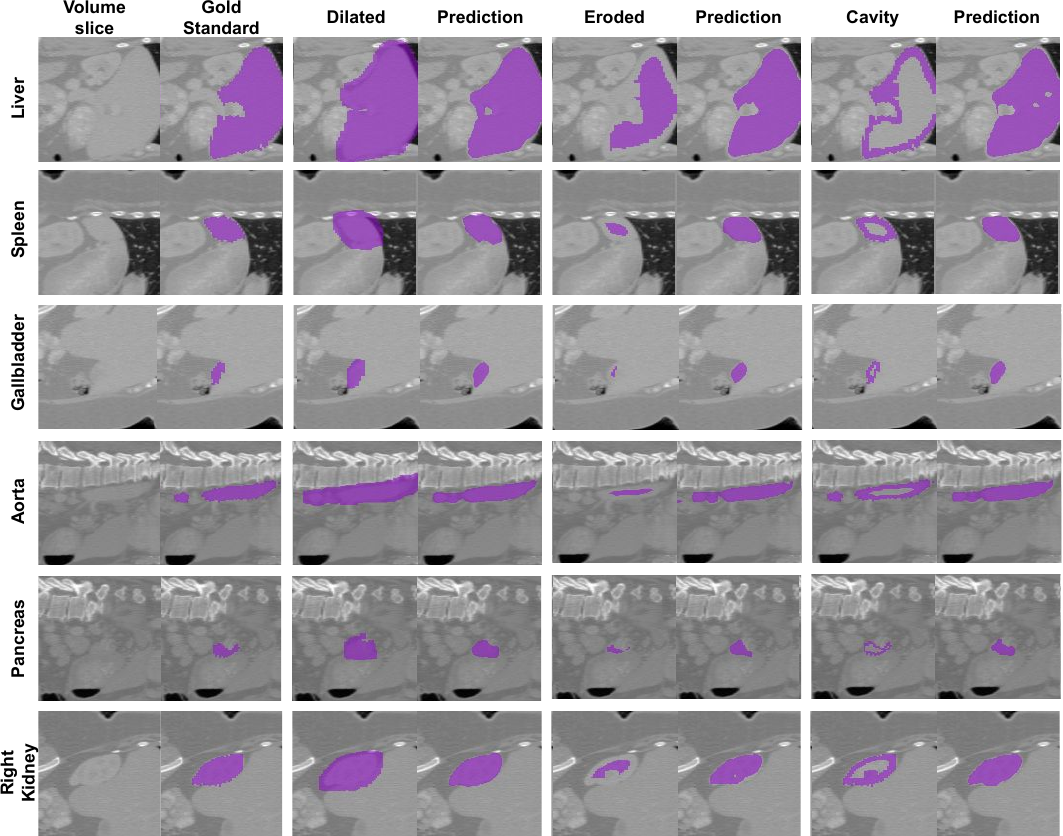}
    \vspace{-0.7cm}
    \caption{Qualitative results of SAM-Med3D for different prompt perturbations. Each row depicts a different organ.}
    \vspace{-0.4cm}
    \label{fig:sam3d_augs}
\end{figure}

\begin{figure*}[t]
    \centering
    \includegraphics[width=1\textwidth]{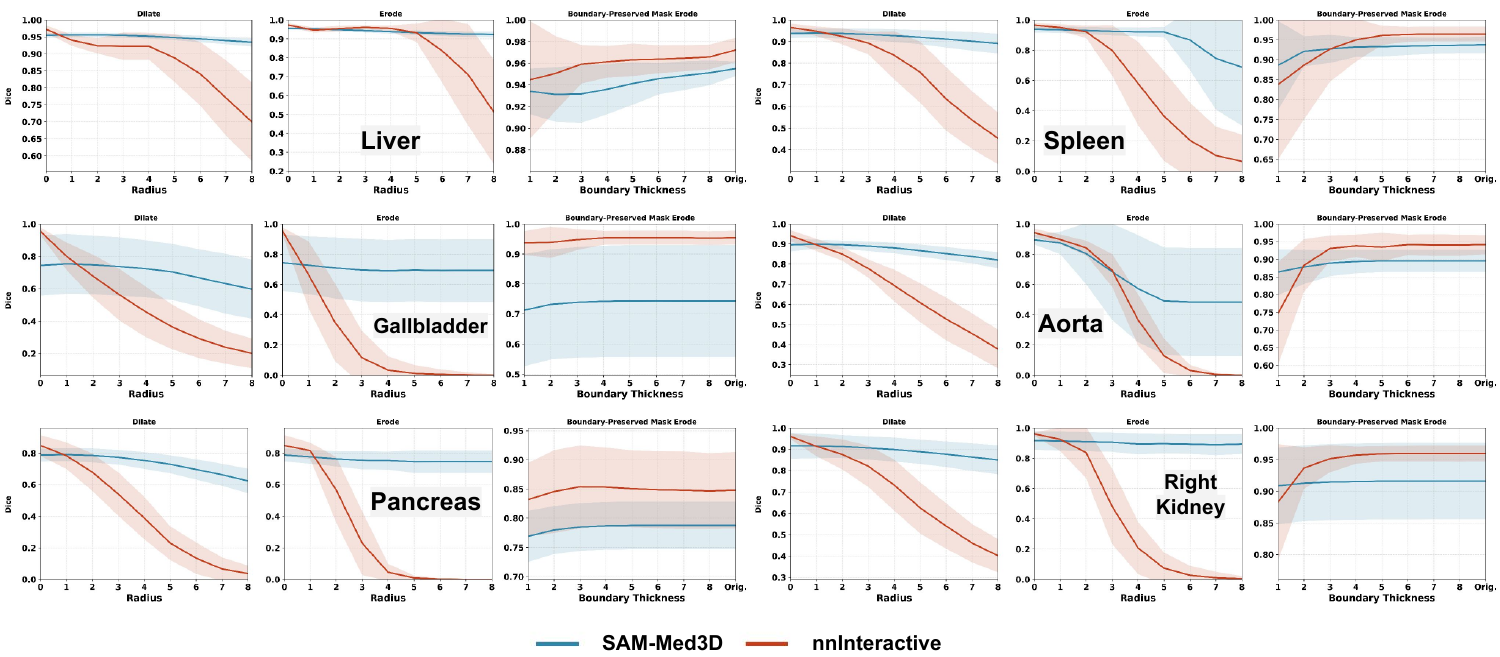}
    \vspace{-0.7cm}
    \caption{Segmentation trends across varying strengths of prompt perturbations (radius for dilation/erosion, boundary thickness for boundary-preserved erosion) for different organs. The respective color bands depict the standard deviation across all samples.}
    \vspace{-0.35cm}
    \label{fig:organwise}
\end{figure*}

We next examine organ-specific trends for each of these prompt perturbations. We choose representatives across the wide range of geometries of abdominal organs in BTCV~\cite{btcv}: \textit{liver} (large); \textit{spleen} and \textit{kidney} (medium-sized), \textit{gallbladder} (small), \textit{pancreas} (irregular-shaped), and \textit{aorta} (tubular shape). The trends are shown in \autoref{fig:organwise}. We observe that eroded prompts are always worse than dilated prompts, although the geometry of the organs drives their resilience against these perturbations. For a bigger structure like liver, the segmentation performance deteriorates at a lower rate (Dice shows very little drop up to radius of dilation/erosion = 5), whereas for spleen, kidney and gallbladder (which are smaller structures), the drop is more aggressive even for smaller radii. Irregular and tubular structures (pancreas and aorta respectively) show greater rates of deterioration with increasing radius of dilation/erosion. As shown in the global trends, models are more robust to boundary-preserved erosion, where the shape and bounds of segmentation are present. The qualitative visualiztions of the prompts and the predicted segmentations in \autoref{fig:nnint_augs} and \autoref{fig:sam3d_augs} further depict how the variation of organ geometry (and in turn, the prompts) degrades the performance of the segmentation in these organs. As for model trends, nnInteractive always yields superior segmentation with more precise prompts, but it degrades rapidly with increasing coarseness compared to SAM-Med3D (more discussion follows in a later section).

\begin{tcolorbox}[highlightbox]
    \textbf{Bigger organs are more robust to small spatial shifts in dense prompts, while smaller organs suffer more.}
\end{tcolorbox}

\begin{figure}
    \centering
    \includegraphics[width=1\linewidth]{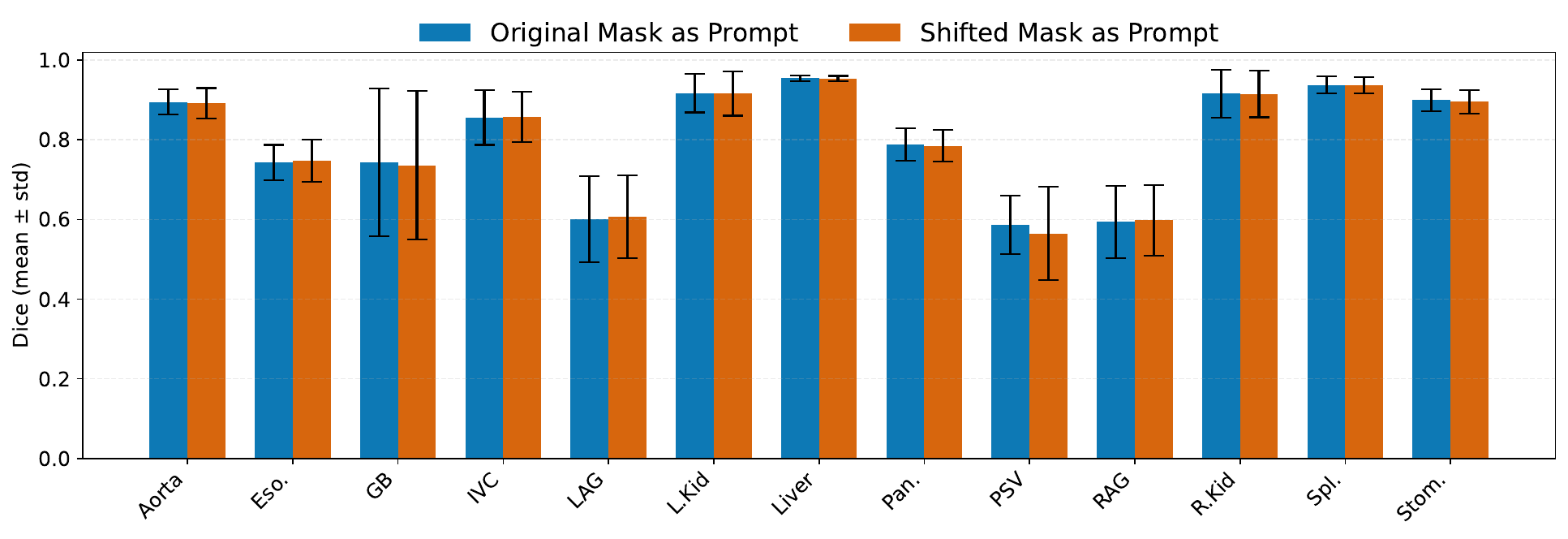}\;\;
    \includegraphics[width=1\linewidth]{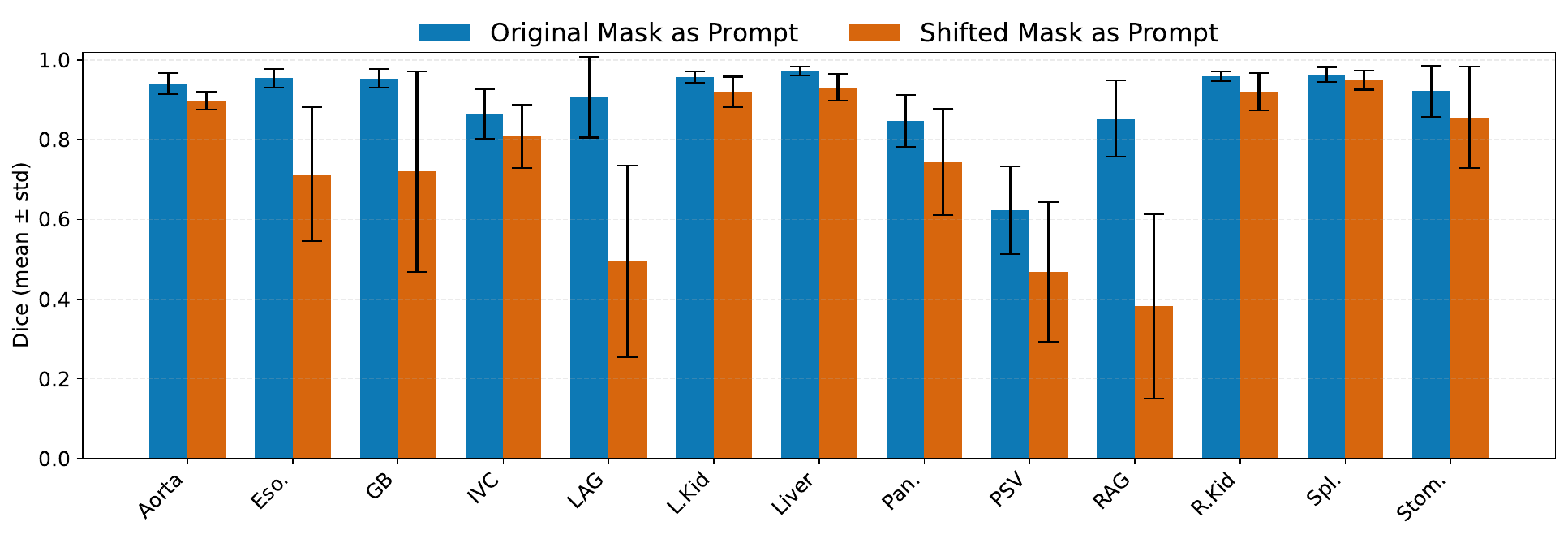}
    \vspace{-0.8cm}
    \caption{Organ-wise performance for \underline{spatially shifted prompts}, for SAM-Med3D \textit{(top)} and nnInteractive \textit{(bottom)}.}
    \vspace{-0.4cm}
    \label{fig:shifts}
\end{figure}

\begin{figure}
    \centering
    \includegraphics[width=1\linewidth]{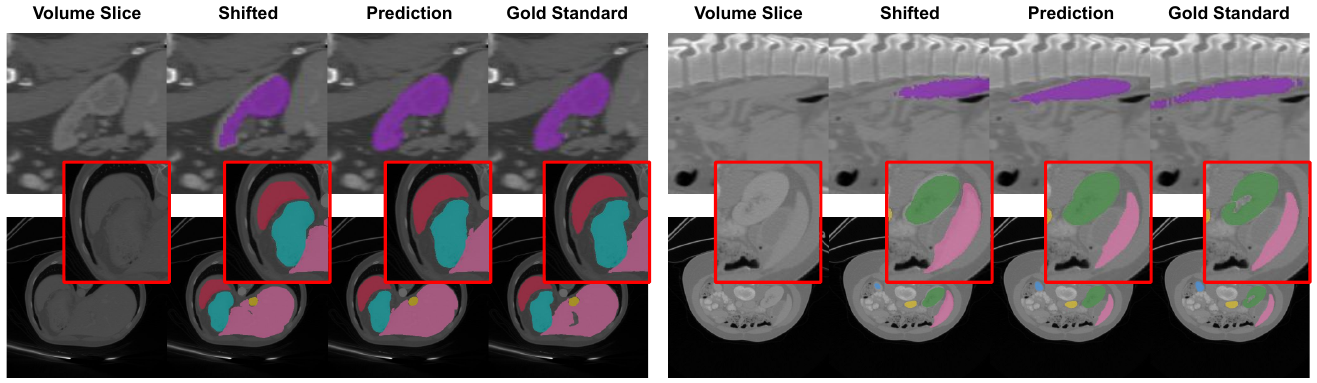}
    \vspace{-0.8cm}
    \caption{Qualitative visuals depicting \underline{spatially shifted prompts} (zoomed-in) and their respective predictions for abdominal organs, using SAM-Med3D \textit{(top)} and nnInteractive \textit{(bottom)}.}
    \vspace{-0.5cm}
    \label{fig:shift_vis}
\end{figure}

We also test the promptable models with spatially misaligned (i.e. shifted) dense prompts for each organ. \autoref{fig:shifts} shows the empirical variations of SAM-Med3D \textit{(top)} and nnInteractive \textit{(bottom)}. We find SAM-Med3D to be robust to such spatial shifts for most organs. For nnInteractive, we observe that segmentation degradations are lower for regular shaped organs such as liver, kidneys and spleen, while more so in smaller organs (adrenal glands, gallbladder) and organs with irregular shape (portal and splenic vein, esophagus). A fundamental difference between spatial translation and morphological operations is that the former preserves the original geometry of the target structure, making the prompt geometry-aware (similar to boundary-preserving erosion), and so the models can better recover the original organ from the spatially shifted input compared to dilation/erosion. We also show qualitative samples in \autoref{fig:shift_vis} for various organs. Combining with earlier observations from boundary-preserved eroded prompts, we hypothesize that \textit{shape-aware visual prompts are \underline{stronger} than mere spatial location-based visual cues} to drive promptable 3D segmentation.

\begin{tcolorbox}[highlightbox]
    \textbf{nnInteractive vs. SAM-Med3D: Broader Discussion}
\end{tcolorbox}

 While we show and analyze trends between the two foundational segmentation models nnInteractive~\cite{nninteractive} and SAM-Med3D~\cite{sammed3d} across perturbed prompts, it is also essential to include a broader discussion about the models themselves. Firstly, nnInteractive injects prompts at original resolution (i.e. in separate channels along with the input image), while SAM-Med3D follows the SAM~\cite{sam} paradigm and injects prompts at downsampled features. Secondly, SAM-Med3D applies a $128^3$ hard cropping and resizing around the target organ, which is fed into the model. While this allows for faster processing, it is also not realistic to expect a user to crop the ROI every time. Secondly, the ROI cropping allows the target region to be a major part the field of view at all times, which may contribute to the relatively higher stability of SAM-Med3D towards prompt perturbations compared to nnInteractive, which contrarily processes all data at native resolution. However, nnInteractive applies a computationally expensive auto-zoom mechanism that adaptively expands the ROI to include the entirety of the organ (please refer to~\cite{nninteractive} for an in-depth understanding), which slows down the overall processing speed in return for higher precision. Thirdly, from \autoref{fig:shifts}, we see that \textit{without any perturbation to the prompts, nnInteractive significantly outperforms SAM-Med3D on all organs.}

\vspace{-0.3cm}
\section{Conclusion}
\vspace{-0.1cm}

We explored the sensitivity of 3D promptable medical segmentation models to varying imprecision in visual prompts. Using controlled synthetic experiments mimicking real-world unavailability of precise visual prompts, we showed existing models generally struggle to overcome perturbed prompts, often over- or under-segmenting target structures, especially for smaller or irregular-shaped organs. Through boundary-preserved erosion simulation, we revealed the usefulness of prominent contextual shape information rather than dense textural cues, which may lead to new directions in training prompt-based segmentation models. In the future, we will leverage these observations and develop robust training paradigms for reliable visual-prompted segmentation models.

\section{Acknowledgements}

This research was, in part, funded by the National Institutes of Health (NIH) under other transactions 1OT2OD038045-01 and NIAMS 1R01AR082684. The views and conclusions contained in this document are those of the authors and should not be interpreted as representing official policies, either expressed or implied, of the NIH.

\section{Compliance with Ethical Standards}



This research study was conducted retrospectively using human subject data made available in open access from the datasets aptly cited in our paper. Ethical approval was not required, as confirmed by the respective licenses attached with the open access data.

\bibliographystyle{IEEEbib}
\bibliography{strings,main}

\end{document}